# Integrated switch for simultaneous mode-division multiplexing (MDM) and wavelength-division multiplexing (WDM)

Brian Stern[1], Xiaoliang Zhu[2], Christine P. Chen[2], Lawrence D. Tzuang[1], Jaime Cardenas[1], Keren Bergman[2] & Michal Lipson[1,3]

[1]School of Electrical and Computer Engineering, Cornell University, Ithaca, NY 14853, USA
[2]Department of Electrical Engineering, Columbia University, New York, NY 10027, USA
[3]Kavli Institute at Cornell for Nanoscale Science, Cornell University, Ithaca, NY 14853, USA
Correspondence and requests for materials should be addressed to M.L. (email: ML292@cornell.edu)

Leveraging the spatial modes of multimode waveguides using mode-division multiplexing (MDM) on an integrated photonic chip allows unprecedented scaling of bandwidth density for on-chip communication. Switching channels between waveguides is critical for future scalable optical networks, but its implementation in multimode waveguides must address how to simultaneously control modes with vastly different optical properties. Here we present a platform for switching signals between multimode waveguides based on individually processing the spatial mode channels using single-mode elements. Using this wavelength-division multiplexing (WDM) compatible platform, we demonstrate a 1x2 multimode switch for a silicon chip which routes four data channels with low (<-20 dB) crosstalk. We show bit-error rates below $10^{-9}$ and power penalties below 1.4 dB on all channels while routing 10 Gbps data when each channel is input and routed separately.  The switch exhibits an additional power penalty of less than 2.4 dB when all four channels are simultaneously routed.

Mode-division multiplexing (MDM) offers a new dimension to scale on-chip bandwidth by utilizing the spatial modes of waveguides to carry multiple optical signals simultaneously[1–21]. The ability to switch and route such channels through a reconfigurable network would enable new functionalities for MDM, which, when combined with wavelength-division multiplexing (WDM), has been projected to allow over 4 Tbps data rate on a single multimode waveguide[18]. However, switching has only been achieved in single-mode on-chip networks[22–27]. The difficulty in implementing switching for multimode waveguides is due to the contradictory design requirements: since the mode confinements in a multimode waveguide vary significantly between the different modes, the dimensions of the photonic structure required to perform the switching differ greatly from mode to mode as well. In fiber communication, despite the fact that spatial multiplexing has allowed enormous data rates over kilometers of fiber[28–39], its small index contrast ($\Delta n \sim 5 \cdot 10^{-3}$) makes coupling between modes rather strong, and therefore modes are not easily separable and switching is confined only to the wavelength domain[40–44]. In integrated silicon waveguides, due to the much higher index contrast ($\Delta n \sim 2$), coupling between modes is much weaker and therefore an integrated multimode platform could allow arbitrary access to individual spatial modes and wavelengths alike to enable reconfigurable switching[45,46] for fully flexible, dense, on-chip optical networks. In this article, we present an integrated multimode switch and demonstrate routing for simultaneous MDM and WDM on-chip. The switch routes four 10 Gbps data channels independently between multimode waveguides with <-20 dB measured crosstalk between modes.

**Results**

We propose a platform for active, integrated multimode photonics based on the independent processing of the spatial modes' signals using single-mode elements. This approach leverages the high index contrast on-chip, which in turn enables access to the individual modes. In the proposed platform, the input multimode signals are first all converted into the fundamental mode, as illustrated in step (1) in Fig. 1 for the example case of 12 channels, consisting of three modes and four wavelengths. Once the modes are converted, processing of the individual channels, now all accessible regardless of mode or wavelength, is possible, including variable attenuation (step (2) in Fig. 1), switching or modulation. Following the processing step, the channels are then reconverted into their original spatial modes at the output (step (3) in Fig. 1).

As an example of the proposed platform we show a multimode 1x2 switch for a silicon chip that supports four data channels, based on ring resonators for switching and for converting the different modes and wavelengths. The switch routes four channels, consisting of two transverse electric modes, $TE_0$ (fundamental) and $TE_1$, at two wavelengths near 1550 nm, from a single input to either of the two output ports (Fig. 2a). Each of the four channels can be routed independently of each other for full switching selectivity. An example switching configuration is shown in Fig. 2b. In order to convert all channels into the fundamental mode and back (stages (1) and (3) in Fig. 2b), we optimize the waveguide widths to ensure phase-matching between the different modes in the waveguides: the $TE_1$ in the 930-nm wide multimode bus waveguide and the $TE_0$ in the 450-nm wide single-mode waveguides (see Supplementary Fig. 1). We utilize racetrack ring resonators to enhance the coupling between these modes for efficient conversion within a short coupling section, as demonstrated in our previous work[18]. The design process for the parameters and dimensions of the waveguides and rings are detailed further in the Methods section. The switching backbone (stage (2) in Fig. 2b) also consists of ring resonators to allow for compact, active control by integrated Ni heaters[47]. These rings have a smaller radius of 8.6 µm (for a free spectral range (FSR) of 10 nm) and are only tuned into resonance when the desired channel is set to be switched, in contrast to the rings employed for mode conversion which have a larger 16 µm radius (for an FSR of 5 nm) and are always kept tuned on resonance so that at all wavelengths the channels are converted between modes[48]. Note that, in principle, increasing the number of rings used for switching and tripling, quadrupling, etc. the FSRs of the rings used for conversion would enable additional wavelength channels.

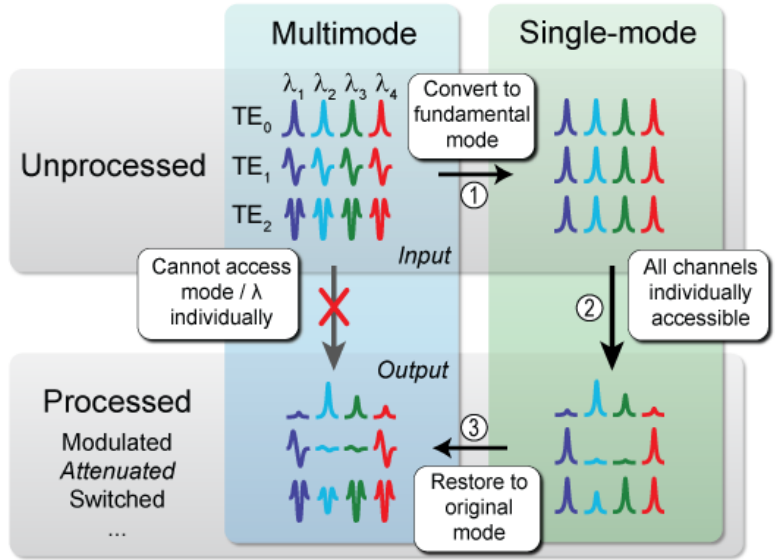

**Figure 1 | Diagram of multimode processing technique.** In order to enable access to individual mode-multiplexed channels, multimode signals are temporarily encoded as the fundamental mode (step (1)), and are then processed independently (step (2)). Processing by variable attenuation is shown as an example for the case of 12 channels (three modes and four wavelengths). Finally, the channels are restored as higher-order spatial modes (step (3)) and coupled to a multimode waveguide output.

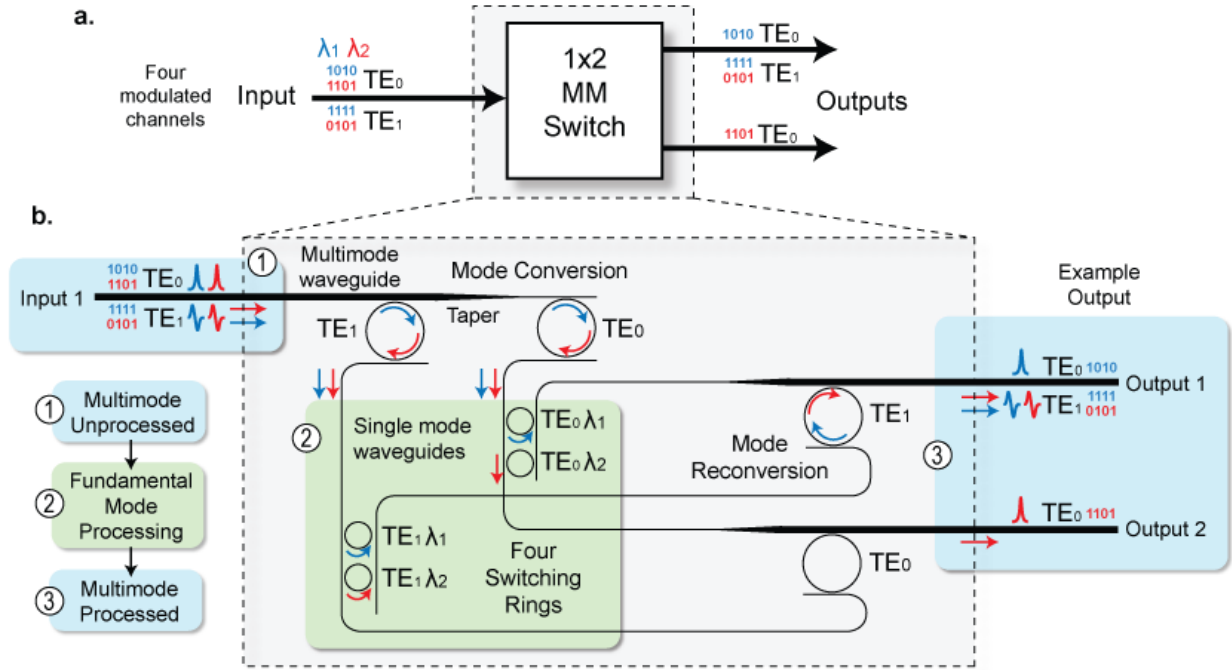

**Figure 2 | Multimode switching design.** (a) Block diagram of 1x2 multimode switch operation. The four input data channels, consisting of two modes at two different wavelengths, may be switched in any combination to the two outputs. The example shows three channels routed to Output 1 and one channel to Output 2. (b) Schematic of multimode switch. The input $TE_1$ channels are converted to the fundamental mode through phase-matching to single-mode rings. The channels are switched using actively-tuned rings to route them individually. This example shows three channels routed from Input 1 to Output 1 using on-resonance rings, and the fourth channel ($TE_0\lambda_2$) is routed from Input 1 to Output 2 using an off-resonance ring.

We measure <-20 dB crosstalk when switching each of the four individual channels to the different output ports. Figure 3a-b shows microscope and scanning electron microscope (SEM) images of the on-chip multimode switch, fabricated on a silicon-on-insulator (SOI) wafer. A detailed explanation of the fabrication process is found in the Methods section. In order to couple on and off the chip using single-mode edge coupling based on an inverted taper[49], a mode (de)multiplexer is added to the input (outputs) of the switch[18]. We measure the intermodal crosstalk between channels by launching one channel at a time and detecting the power at each output to compare the desired signal with leaked, interfering signals from other channels (Fig. 3c). For all channels and switching configurations, the crosstalk is less than -20 dB, ranging from -28.5 to -20.1 dB for each channel. These low crosstalk values are comparable to previous integrated multimode multiplexer systems[10,13,18], indicating that the switch introduces negligible crosstalk. The measured insertion loss, including on- and off-chip coupling losses, is 5.4 to 9.1 dB for the different four channels. Based on the measured losses from test structures that do not include the switch or multiplexers, we estimate the losses due to the switch and multiplexers together to be between 0.9 dB and 4.6 dB This mode-dependent insertion loss is mainly due to variations in the rings' coupling and therefore extinction ratio[50], which leads to greater losses for the higher-order modes in this case. These fabrication-induced variations can in principle be overcome using tunable couplers such as interferometers[51].

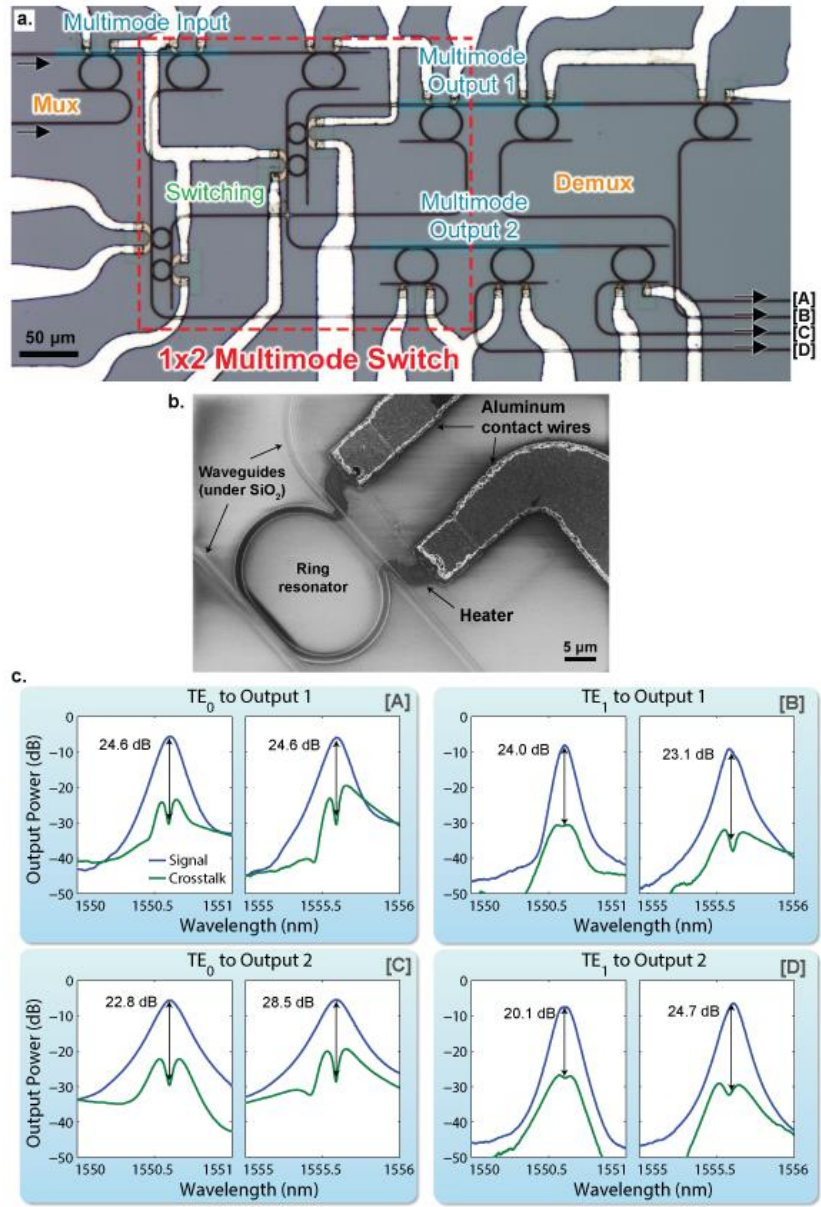

**Figure 3 | Fabricated device and crosstalk measurements.** (a) Optical microscope image of fabricated device. The input channels are coupled into single-mode waveguides from an off-chip laser, and a multiplexer (mux) produces the MDM input to the multimode switch. The areas highlighted in blue show the multimode waveguides. The four small rings actively switch the four channels and are tuned by integrated heaters. Following the switch, each of the two outputs is demultiplexed (demux) so that the channels can be individually monitored off-chip. Scale bar is 50 μm. (b) SEM image of a ring resonator in the fabricated device with active heater used to tune the resonance. Scale bar is 5 μm. (c) Crosstalk

measurements for the different channels. Spectral profiles at both outputs for each of the input four channels, compared with the profiles from interfering channels. Signal and crosstalk were measured individually with a continuous wave tunable laser for the configurations with highest intermodal crosstalk, which remain below -20 dB in all cases.

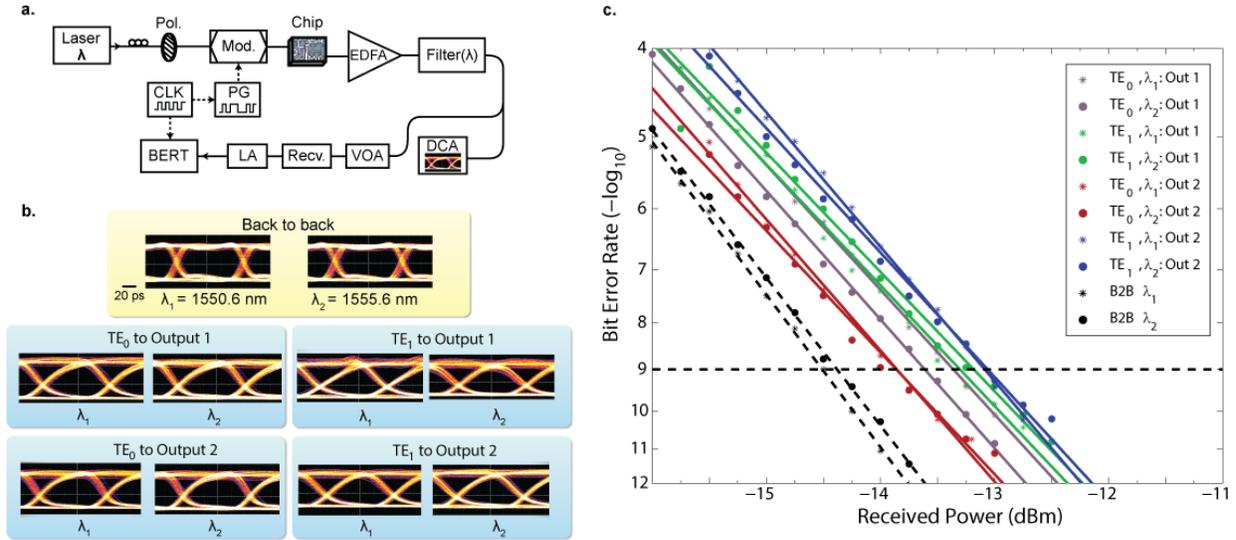

**Figure 4 | Error-free switching of 10 Gbps MDM-WDM data.** (a) Testing configuration, including tunable laser, polarization controller, fiber polarizer, electro-optic modulator (Mod.), pattern generator (PG) for the $2^7-1$ pseudo-random binary sequence (PRBS), function generator clock source (CLK), Er-doped fiber amplifier (EDFA), tunable band-pass filter (1.4 nm), digital communications analyzer (DCA), variable optical attenuator (VOA), optical receiver (Recv.), limiting amplifier (LA), and bit-error rate tester (BERT). (b) Eye diagrams of the switched signals for all channels at both outputs are open. Comparison with the rise time of back-to-back eyes confirms that the output signal is bandwidth-limited. Scale bar (first eye) is 20 ps. (c) Error free transmission (BER < $10^{-9}$) is achieved with power penalties ranging from 0.5 to 1.4 dB, compared to the back to back (B2B) references.

Our 1x2 multimode switch exhibits bit-error rates (BER) below $10^{-9}$ on all channels, and open eye diagrams while routing 10 Gbps data when each channel is input and routed separately. We perform the experiment using a tunable laser modulated by a pseudo-random binary sequence (PRBS) from a pattern generator (Fig. 4a). The modulated light is coupled onto the chip using a tapered fiber. A DC voltage is applied to each integrated heater to align their resonances with the laser or to tune and detune the resonances of the rings used for switching to route the channels to the outputs. The total power supplied to the heaters is up to 30 mW, depending on the switching state, and is almost entirely used for aligning the resonances of all rings due to fabrication variations. A back-to-back reference for the transmission experiment is measured for each wavelength by removing the chip and replacing the tapered fibers with a single fiber connection. The output signal from the chip is amplifed and filtered (to reject amplified spontaneous emission (ASE) noise) to obtain optical eye diagrams of the transmitted data pattern (Fig. 4b). One can see that the signals exhibit open eye diagrams for all four channels routed to either output. We further characterize the data integrity with BER measurements (Fig. 4c). We measure error-free switching (BER < 10$^{-9}$) for all channels, with the power penalty ranging between 0.5 and 1.0 dB for TE$_0$ and 1.2 to 1.4 dB for TE$_1$.

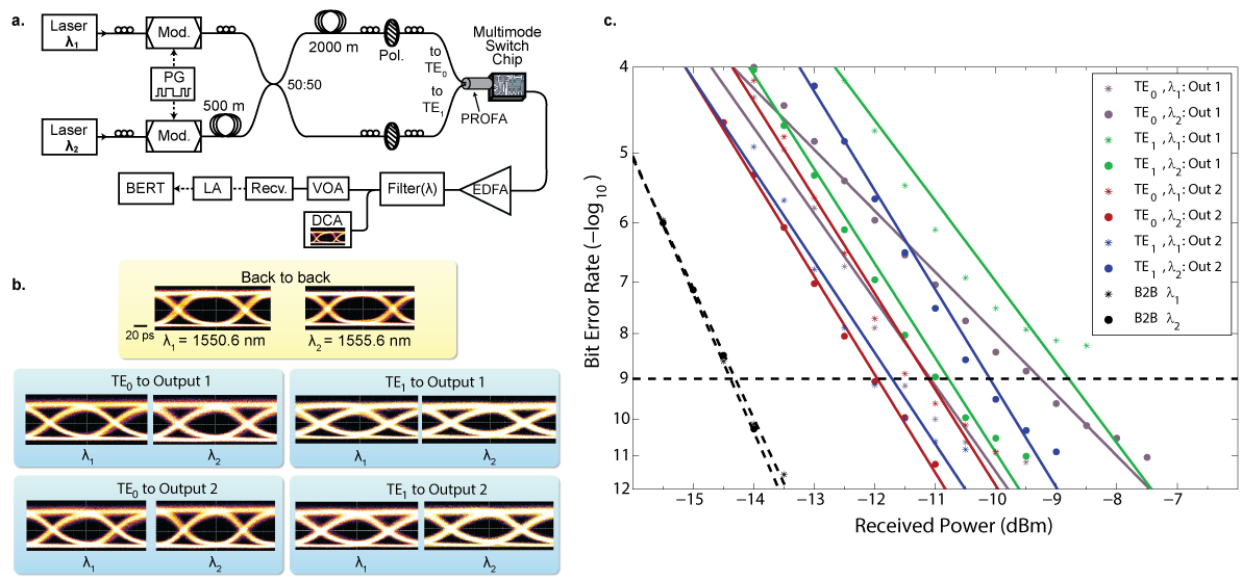

**Figure 5 | Simultaneous operation of four WDM-MDM channels.** (a) Testing setup for simultaneous switching, derived from that of Fig. 4a, includes fiber spools for decorrelation and decoherence of data channels, a 50:50 combiner/splitter, and pitch reducing optical fiber array (PROFA). Solid connections are in fiber, and dotted connections are electrical. (b) Eye diagrams for each 10 Gbps channel switching to either output. Scale bar (first eye) is 20 ps. (c) Bit-error rate measurements for simultaneous operation of all channels. The worst-case switching configuration for each channel is plotted. For each channel, a back to back reference was measured by replacing the PROFA and chip with an attenuator replicating the insertion loss. The best-performing back to back measurements for each wavelength are plotted. All channels achieve error-free (BER < 10$^{-9}$) transmission, except the TE$_1\lambda_1$ channel, which is impaired when switched to Output 1 due to fabrication error causing one ring resonator to be under-coupled with a narrow bandwidth.

The switch exhibits an additional power penalty of less than 2.4 dB when all four channels are simultaneously inserted onto the chip and routed. In order to accommodate two wavelength channels ($\lambda_1$ and $\lambda_2$), we decorrelate them using 500 m of single-mode fiber (SMF) and then combine and split them equally on two paths (Fig. 5a). Another length of 2 km of fiber on one path ensures phase decoherence of the two paths, as it is several times the laser coherence length of 450 m. Each path has equal power and is coupled simultaneously into the TE$_0$ and TE$_1$ inputs respectively using a pitch reducing optical fiber array (PROFA)[52]. A tapered fiber is used to selectively measure the outputs of the chip by mode, and the tunable filter is aligned to measure by wavelength channel. We observe open eye diagrams for the four simultaneously routed 10 Gbps channels (Fig. 5b). The back to back references are measured by replacing the chip and PROFA with an attenuator to replicate each path's insertion loss. We also measure the BER for all channels (Fig. 5c) for the highest crosstalk configuration, i.e. when the mode channels are routed to the same output. We observe power penalties of 2.4 to 5.1 dB for simultaneous operation, and error-free switching (BER < $10^{-9}$) for all channels except TE$_1\lambda_1$ at Output 1, which reaches $5\cdot10^{-9}$. The additional power penalty of 1.8 to 2.4 dB due to simultaneous operation could be minimized by optimizing the bandwidth of each ring to equalize the effect of intrachannel crosstalk among the different paths. The higher power penalty for channel TE$_1\lambda_1$ at Output 1 is due to fabrication error causing one switching ring to be under-coupled, resulting in a narrow bandwidth of only 9 GHz, compared to the larger bandwidth for the other channels of approximately 13 GHz.

**Discussion**

This demonstration of the first integrated multimode switch establishes MDM as a viable platform for optical interconnects. It allows scaling of bandwidth density for on-chip networks by expanding routing to include waveguides employing simultaneous MDM and WDM for multi-dimensional multiplexing. The ability to route on-chip MDM-WDM signals with full flexibility enables integrated networks with many nodes connected by high-bandwidth multimode links to dynamically allocate bandwidth. While each multimode input or output in this demonstration carries 40 Gbps aggregate bandwidth (4 x 10 Gbps), the design is scalable in principle to more modes and wavelengths. The platform we present for processing multimode signals in the single-mode domain also creates the possibility for numerous future applications of MDM beyond routing, such as modulation, attenuation, or performance monitoring.

**Methods**

**Phase matching waveguide widths.** Asymmetric coupling regions are used in the multiplexing, demultiplexing, and mode conversion steps of the switch. The multimode waveguides are designed to be 930 nm wide to accommodate phase-matching of the TE$_1$ mode with the TE$_0$ mode of 450-nm wide single-mode waveguides (see Supplementary Fig. 1)[18]. At this combination of widths, the effective indices of the modes in their respective waveguides match with index 2.46, and so the TE$_0$ mode in the single-mode waveguide selectively excites the TE$_1$ mode in the multimode waveguide, without also exciting the TE$_0$ therein, which would contribute to crosstalk. If the fundamental mode alone is present in a waveguide, it can be adiabatically tapered wider or narrower without disturbing the mode.

**Coupling design.** The racetrack ring resonators used for the (de)multiplexers have radii of 16 μm and coupling lengths of 5.9 μm, at which the crosstalk is minimized due to the phase-matching. The switching rings have 8.6 μm radii and 1.2 μm coupling lengths, for half the circumference of the multiplexing rings. The coupling gaps between the rings and waveguides are chosen to meet the critical coupling condition κ = κ' + α, where κ and κ' are the add and drop port coupling constants, respectively[50]. To enable 10 Gbps operation, κ and κ' are also optimized for a bandwidth of 16 GHz per ring, although this narrows when rings are in series. The coupling gaps are listed in Supplementary Table 1. The adiabatic taper length is 95 μm.

**Device fabrication.** We fabricate the switch on a silicon-on-insulator (SOI) wafer with 250-nm thick Si device layer on 3 μm buried oxide. The waveguides are patterned using electron beam lithography and fully etched using reactive ion etching. The devices are then clad with 1 μm of plasma enhanced chemical vapor deposition (PECVD) SiO$_2$. A thin Cr adhesion layer and 100 nm of Ni are evaporated along with a lift-off process to define the heaters for tuning resonances. For the metal contacts, 1.7 um of Al is sputtered with a thin Ti adhesion layer and then etched using inductively coupled plasma. Deep trenches are etched into the silicon substrate near the input and output waveguide tapers for improved coupling[49,53]. The final chip is mounted to a custom printed circuit board (PCB), onto which the Al pads are wirebonded out for easy control of heater tuning by DC voltages. The switch area is <0.07 mm$^2$, and an even more compact design could be achieved by using smaller tapers or placing components closer together.

**References**

1. Lee, B.-T. & Shin, S.-Y. Mode-order converter in a multimode waveguide. *Opt. Lett.* **28,** 1660–1662 (2003).
2. Greenberg, M. & Orenstein, M. Multimode add-drop multiplexing by adiabatic linearly tapered coupling. *Opt. Express* **13,** 9381–9387 (2005).
3. Bagheri, S. & Green, W. M. J. Silicon-on-insulator mode-selective add-drop unit for on-chip mode-division multiplexing. in *6th IEEE International Conference on Group IV Photonics, 2009. GFP '09* 166–168 (2009). doi:10.1109/GROUP4.2009.5338328
4. Ruege, A. C. & Reano, R. M. Multimode Waveguides Coupled to Single Mode Ring Resonators. *J. Light. Technol.* **27,** 2035–2043 (2009).
5. Frandsen, L. H. *et al.* Topology optimized mode conversion in a photonic crystal waveguide fabricated in silicon-on-insulator material. *Opt. Express* **22,** 8525–8532 (2014).
6. Heinrich, M. *et al.* Supersymmetric mode converters. *Nat. Commun.* **5,** (2014).
7. Hanzawa, N. *et al.* Mode multi/demultiplexing with parallel waveguide for mode division multiplexed transmission. *Opt. Express* **22,** 29321–29330 (2014).
8. Chen, W., Wang, P. & Yang, J. Mode multi/demultiplexer based on cascaded asymmetric Y-junctions. *Opt. Express* **21,** 25113–25119 (2013).
9. Chen, C. P. *et al.* Experimental Demonstration of Spatial Scaling for High-Throughput Transmission Through A Si Mode-Division-Multiplexing Waveguide. in *Advanced Photonics for Communications* IM2A.3 (Optical Society of America, 2014). doi:10.1364/IPRSN.2014.IM2A.3
10. Driscoll, J. B. *et al.* Asymmetric Y junctions in silicon waveguides for on-chip mode-division multiplexing. *Opt. Lett.* **38,** 1854–1856 (2013).
11. Driscoll, J. B. *et al.* A 60 Gb/s MDM-WDM Si photonic link with < 0.7 dB power penalty per channel. *Opt. Express* **22,** 18543–18555 (2014).
12. Uematsu, T., Ishizaka, Y., Kawaguchi, Y., Saitoh, K. & Koshiba, M. Design of a Compact Two-Mode Multi/Demultiplexer Consisting of Multimode Interference Waveguides and a Wavelength-Insensitive Phase Shifter for Mode-Division Multiplexing Transmission. *J. Light. Technol.* **30,** 2421–2426 (2012).
13. Wang, J., Chen, P., Chen, S., Shi, Y. & Dai, D. Improved 8-channel silicon mode demultiplexer with grating polarizers. *Opt. Express* **22,** 12799 (2014).
14. Wang, J., Chen, S., Chen, P., Shi, Y. & Dai, D. 64-channel hybrid (de)multiplexer enabling wavelength- and mode-division multiplexing for on-chip optical interconnects. in *Asia Communications and Photonics Conference 2014* ATh1A.7 (Optical Society of America, 2014). doi:10.1364/ACPC.2014.ATh1A.7
15. Yin, M., Deng, Q., Li, Y., Wang, X. & Li, H. Compact and broadband mode multiplexer and demultiplexer based on asymmetric plasmonic–dielectric coupling. *Appl. Opt.* **53,** 6175–6180 (2014).
16. Ding, Y. *et al.* On-chip two-mode division multiplexing using tapered directional coupler-based mode multiplexer and demultiplexer. *Opt. Express* **21,** 10376–10382 (2013).

17. Yang, Y.-D., Li, Y., Huang, Y.-Z. & Poon, A. W. Silicon nitride three-mode division multiplexing and wavelength-division multiplexing using asymmetrical directional couplers and microring resonators. *Opt. Express* **22,** 22172–22183 (2014).
18. Luo, L.-W. *et al.* WDM-compatible mode-division multiplexing on a silicon chip. *Nat. Commun.* **5,** (2014).
19. Mulugeta, T. & Rasras, M. Silicon hybrid (de)multiplexer enabling simultaneous mode and wavelength-division multiplexing. *Opt. Express* **23,** 943–949 (2015).
20. Dorin, B. A. & Ye, W. N. Two-mode division multiplexing in a silicon-on-insulator ring resonator. *Opt. Express* **22,** 4547–4558 (2014).
21. Chen, H., van Uden, R., Okonkwo, C. & Koonen, T. Compact spatial multiplexers for mode division multiplexing. *Opt. Express* **22,** 31582–31594 (2014).
22. Almeida, V. R., Barrios, C. A., Panepucci, R. R. & Lipson, M. All-optical control of light on a silicon chip. *Nature* **431,** 1081–1084 (2004).
23. Dong, P., Preble, S. F. & Lipson, M. All-optical compact silicon comb switch. *Opt. Express* **15,** 9600–9605 (2007).
24. Yang, M. *et al.* Non-Blocking 4x4 Electro-Optic Silicon Switch for On-Chip Photonic Networks. *Opt. Express* **19,** 47–54 (2011).
25. Ji, R. *et al.* Five-port optical router for photonic networks-on-chip. *Opt. Express* **19,** 20258–20268 (2011).
26. Biberman, A. *et al.* Photonic Network-on-chip Architectures Using Multilayer Deposited Silicon Materials for High-performance Chip Multiprocessors. *J Emerg Technol Comput Syst* **7,** 7:1–7:25 (2011).
27. Lee, B. G. *et al.* Monolithic Silicon Integration of Scaled Photonic Switch Fabrics, CMOS Logic, and Device Driver Circuits. *J. Light. Technol.* **32,** 743–751 (2014).
28. Richardson, D. J., Fini, J. M. & Nelson, L. E. Space-division multiplexing in optical fibres. *Nat. Photonics* **7,** 354–362 (2013).
29. Winzer, P. J. Making spatial multiplexing a reality. *Nat. Photonics* **8,** 345–348 (2014).
30. Igarashi, K. *et al.* Super-Nyquist-WDM transmission over 7,326-km seven-core fiber with capacity-distance product of 1.03 Exabit/s·km. *Opt. Express* **22,** 1220–1228 (2014).
31. Sakaguchi, J. *et al.* 305 Tb/s Space Division Multiplexed Transmission Using Homogeneous 19-Core Fiber. *J. Light. Technol.* **31,** 554–562 (2013).
32. Van Uden, R. G. H. *et al.* Ultra-high-density spatial division multiplexing with a few-mode multicore fibre. *Nat. Photonics* **8,** 865–870 (2014).
33. Amaya, N. *et al.* Fully-elastic multi-granular network with space/frequency/time switching using multi-core fibres and programmable optical nodes. *Opt. Express* **21,** 8865 (2013).
34. Takara, H. *et al.* 1.01-Pb/s (12 SDM/222 WDM/456 Gb/s) Crosstalk-managed Transmission with 91.4-b/s/Hz Aggregate Spectral Efficiency. in *European Conference and Exhibition on Optical Communication* Th.3.C.1 (Optical Society of America, 2012). doi:10.1364/ECEOC.2012.Th.3.C.1
35. Fontaine, N. K. *et al.* Space-division multiplexing and all-optical MIMO demultiplexing using a photonic integrated circuit. in *Optical Fiber Communication Conference* PDP5B.1 (Optical Society of America, 2012). doi:10.1364/OFC.2012.PDP5B.1
36. Randel, S. *et al.* 6×56-Gb/s mode-division multiplexed transmission over 33-km few-mode fiber enabled by 6×6 MIMO equalization. *Opt. Express* **19,** 16697 (2011).
37. Ip, E. *et al.* Few-mode fiber transmission with in-line few-mode erbium-doped fiber amplifier. in **8647,** 864709–864709–10 (2013).
38. Ryf, R. *et al.* Mode-Division Multiplexing Over 96 km of Few-Mode Fiber Using Coherent 6 6 MIMO Processing. *J. Light. Technol.* **30,** 521–531 (2012).

39. Ryf, R. *et al.* Mode-multiplexed transmission over conventional graded-index multimode fibers. *Opt. Express* **23,** 235–246 (2015).
40. Ho, K.-P., Kahn, J. M. & Wilde, J. P. Wavelength-Selective Switches for Mode-Division Multiplexing: Scaling and Performance Analysis. *J. Light. Technol.* **32,** 3724–3735 (2014).
41. Ryf, R. *et al.* Wavelength-selective switch for few-mode fiber transmission. in *39th European Conference and Exhibition on Optical Communication (ECOC 2013)* 1–3 (2013). doi:10.1049/cp.2013.1681
42. Chen, X., Li, A., Ye, J., Al Amin, A. & Shieh, W. Demonstration of Few-Mode Compatible Optical Add/Drop Multiplexer for Mode-Division Multiplexed Superchannel. *J. Light. Technol.* **31,** 641–647 (2013).
43. Gu, R. Y., Ip, E., Li, M.-J., Huang, Y.-K. & Kahn, J. M. Experimental Demonstration of a Spatial Light Modulator Few-Mode Fiber Switch for Space-Division Multiplexing. in *Frontiers in Optics 2013 Postdeadline* (ed. Kang, I., Reitze, D., Alic, N., and Hagan, D.) FW6B.4 (Optical Society of America, 2013). doi:10.1364/FIO.2013.FW6B.4
44. Carpenter, J. A. *et al.* 1x11 Few-Mode Fiber Wavelength Selective Switch using Photonic Lanterns. in *Optical Fiber Communication Conference* Th4A.2 (Optical Society of America, 2014). doi:10.1364/OFC.2014.Th4A.2
45. Cvijetic, M., Djordjevic, I. B. & Cvijetic, N. Dynamic multidimensional optical networking based on spatial and spectral processing. *Opt. Express* **20,** 9144–9150 (2012).
46. Dai, D., Wang, J. & He, S. Silicon Multimode Photonic Integrated Devices For On-Chip Mode-Division-Multiplexed Optical Interconnects. *Prog. Electromagn. Res.* **143,** 773–819 (2013).
47. Sherwood-Droz, N. *et al.* Optical 4x4 hitless slicon router for optical networks-on-chip (NoC). *Opt. Express* **16,** 15915–15922 (2008).
48. Chan, J. *et al.* Data Transmission Using Wavelength-Selective Spatial Routing for Photonic Interconnection Networks. in *Optical Fiber Communication Conference/National Fiber Optic Engineers Conference 2011* OThQ3 (Optical Society of America, 2011). doi:10.1364/OFC.2011.OThQ3
49. Almeida, V. R., Panepucci, R. R. & Lipson, M. Nanotaper for compact mode conversion. *Opt. Lett.* **28,** 1302–1304 (2003).
50. Yariv, A. Universal relations for coupling of optical power between microresonators and dielectric waveguides. *Electron. Lett.* **36,** 321–322 (2000).
51. Luo, L.-W. *et al.* High bandwidth on-chip silicon photonic interleaver. *Opt. Express* **18,** 23079 (2010).
52. Kopp, V. I. *et al.* Pitch Reducing Optical Fiber Array for dense optical interconnect. in *2012 IEEE Avionics, Fiber- Optics and Photonics Technology Conference (AVFOP)* 48–49 (2012). doi:10.1109/AVFOP.2012.6344072
53. Cardenas, J. *et al.* High Coupling Efficiency Etched Facet Tapers in Silicon Waveguides. *IEEE Photonics Technol. Lett.* **26,** 2380–2382 (2014).

**Acknowledgements**

This work was performed in part at the Cornell NanoScale Facility, a member of the National Nanotechnology Infrastructure Network, which is supported by the National Science Foundation (NSF) through grant ECCS-0335765, and made use of the Cornell Center for Materials Research Shared Facilities which are supported through the NSF MRSEC program (DMR-1120296). This work was also supported by the NSF through CIAN ERC under grant EEC-0812072. The authors also acknowledge the NSF Graduate Research Fellowship Program and Intel SRC Ph.D. Fellowship.

**Author contributions**

B.S. designed and simulated the device with L.D.T. and J.C. assisting. B.S., X.Z. and C.P.C. conducted the experiments. B.S., L.D.T. and J.C. fabricated the device. M.L. and K.B. supervised the project. B.S. prepared the manuscript. M.L. and K.B. edited the manuscript with input from all co-authors.

**Supplementary**

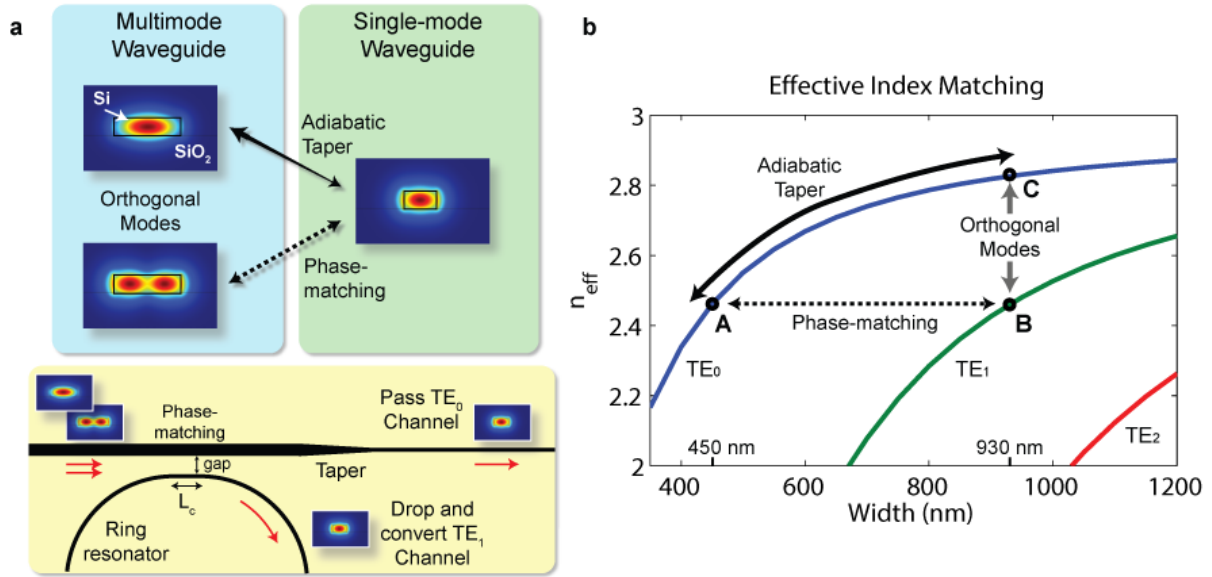

**Supplementary Figure 1 |** (a) Method for accessing individual modes in a multimode waveguide. A single-mode ring resonator which is properly phase-matched to the bus waveguide will drop the $TE_1$ channel, while the $TE_0$ channel can be adiabatically tapered into a single-mode waveguide. (b) Simulated effective index of 250-nm tall Si waveguide as a function of width. The phase-matching condition is met when the index for $TE_0$ in the single-mode waveguide (point A) matches that of $TE_1$ in the multimode bus waveguide (point B). The $TE_0$ signals can adiabatically taper to the same width as the multimode bus waveguide (point C) and back.

| Coupling Region | Gap (nm) |
|---|---|
| $TE_1$ MM to SM ring – Add port | 210 |
| $TE_1$ MM to SM ring – Drop port | 220 |
| $TE_0$ SM to SM ring – Add port | 250 |
| $TE_0$ SM to SM ring – Drop port | 260 |
| Switching Ring – Add port | 190 |
| Switching Ring – Drop port | 200 |

**Supplementary Table 1.** Coupling gaps between waveguides (multimode and single-mode) and ring resonators. The gaps of the conversion rings used in (de)multiplexers and the switching rings are optimized for critical coupling and bandwidth using the eigenmode expansion method to determine the coupling for particular dimensions.